\documentclass[letterpaper]{article} 
\usepackage[preprint]{aaai2027}  
\usepackage[hyphens]{url}  
\usepackage{graphicx} 
\usepackage{natbib}  
\usepackage{caption} 
\usepackage{booktabs}
\usepackage{xcolor}
\usepackage{mdframed}
\newif\ifanonymous
\anonymousfalse

\title{A Corpus of Real Scam- and Spam-Call Conversations\\ from an Active Voice-Agent Honeypot}

\ifanonymous
\author{Anonymous Submission}
\affiliations{}
\else
\author{Ethan Traister, Dennis Tsang Ng, Siyu Zhang, Huaiyu Guo, Tommy Duong,\\
Tyler Wu, Yuchen Zhou, Xingyu Shen, Jiaqi Wu, Simiao Ren\textsuperscript{\dag}}
\affiliations{scam.ai\\ \textsuperscript{\dag}Corresponding author: benren@scam.ai}
\fi

\begin{document}
\maketitle
{\renewcommand{\thefootnote}{}\footnote{Authorship of this paper does not imply employment by or affiliation with Scam.AI (Reality Inc.); some authors are independent researchers who collaborated on this work.}\addtocounter{footnote}{-1}}

\begin{abstract}
Real conversations between fraudsters and their targets are among the most informative artifacts for studying telephone scams, yet they are also among the scarcest: passive telephony honeypots overwhelmingly capture automated messages and immediate hang-ups, large-scale studies characterize call metadata rather than dialogue, and manual scam-baiting does not scale. We present a dataset of real scam-call conversations collected by an \emph{active voice-agent honeypot}. Dedicated telephone numbers are seeded into the lead-generation channels that fraud operations harvest; inbound callers are answered by a low-latency conversational agent that adopts a plausible target persona and sustains the interaction while every call is recorded, transcribed, and automatically labeled. Over an initial 53-day window the system captured \textbf{10,015 inbound scam and spam calls} (6,601 with two or more turns), comprising roughly 895 hours of audio and 328,869 transcribed turns from 5,665 distinct originating numbers, and continues to grow. Under a holistic classifier the substantive calls are predominantly predatory-but-legal lead generation (``spam'', about three in five), while roughly one in seven is an outright ``scam'' (949 in this snapshot)---a fraud attempt through impersonation, credential solicitation, or a payment demand. Each call carries a turn-level transcript, three-channel audio, per-turn latency telemetry, and multiple layers of automatic labels, including a holistic scam/spam/legitimate judgment corroborated by independent human review (75\% agreement on the binary decision). We describe the collection system, the record structure, and technical validation of the corpus's realism and label quality---including that the agent is recognized as non-human in only about 5\% of engaged calls---and benchmark established scam-detection methods on the corpus, where detectors trained on published synthetic dialogue collapse in precision on real traffic. The corpus supports research on fraud detection, social-engineering tactics, and defensive conversational agents.
\end{abstract}

\section{Introduction}
Telephone fraud imposes large and recurring costs on the public. In 2024, U.S.\ consumers reported losing more than \$12.5 billion to fraud, and among the contact methods people reported, the telephone carried the highest median loss---\$1,500 per report \citep{ftc}; the FBI's Internet Crime Complaint Center separately logged \$16.6 billion in reported losses across 859,532 complaints that year \citep{ic3}. Telephone spam alone has been estimated to cost U.S.\ consumers on the order of \$8.6 billion annually \citep{sok,callme}. These threats are increasingly framed as a downside of conversational AI itself: voice-enabled agents can now autonomously place fraudulent calls at negligible cost \citep{aienabledscams,scamagents}. We take the opposite view, and demonstrate that the same capabilities can be turned toward defense---using AI agents not to deceive but to detect, measure, and disrupt fraud at scale. This continues a growing line of work that benchmarks AI-based detectors of AI-enabled fraud and assembles real, in-the-wild datasets of machine-generated content \citep{renbench,gptwild}, which we extend from images and documents to the voice channel. Realizing that promise requires data, yet the artifact that most directly reveals how a scam actually works---the conversation in which an operator delivers a pretext, applies pressure, and solicits money or credentials---remains difficult to obtain at scale.

Existing data-collection strategies each capture only part of the picture. Passive telephony honeypots log inbound traffic but predominantly capture robocalls, silence, and rapid hang-ups rather than sustained dialogue \citep{phoneypot,robocalls}; manual scam-baiting shows that engagement itself is valuable but does not scale, and fixed-script audio bots do not respond to the caller \citep{lenny}; and recent systems that apply language models to engage scammers work chiefly in the email and text-message setting \citep{sendaccount,aiintheloop}. What is still missing is a scalable source of complete, real, consistently labeled scam \emph{conversations} over the voice channel.

\begin{figure}[t]
\centering
\includegraphics[width=\columnwidth]{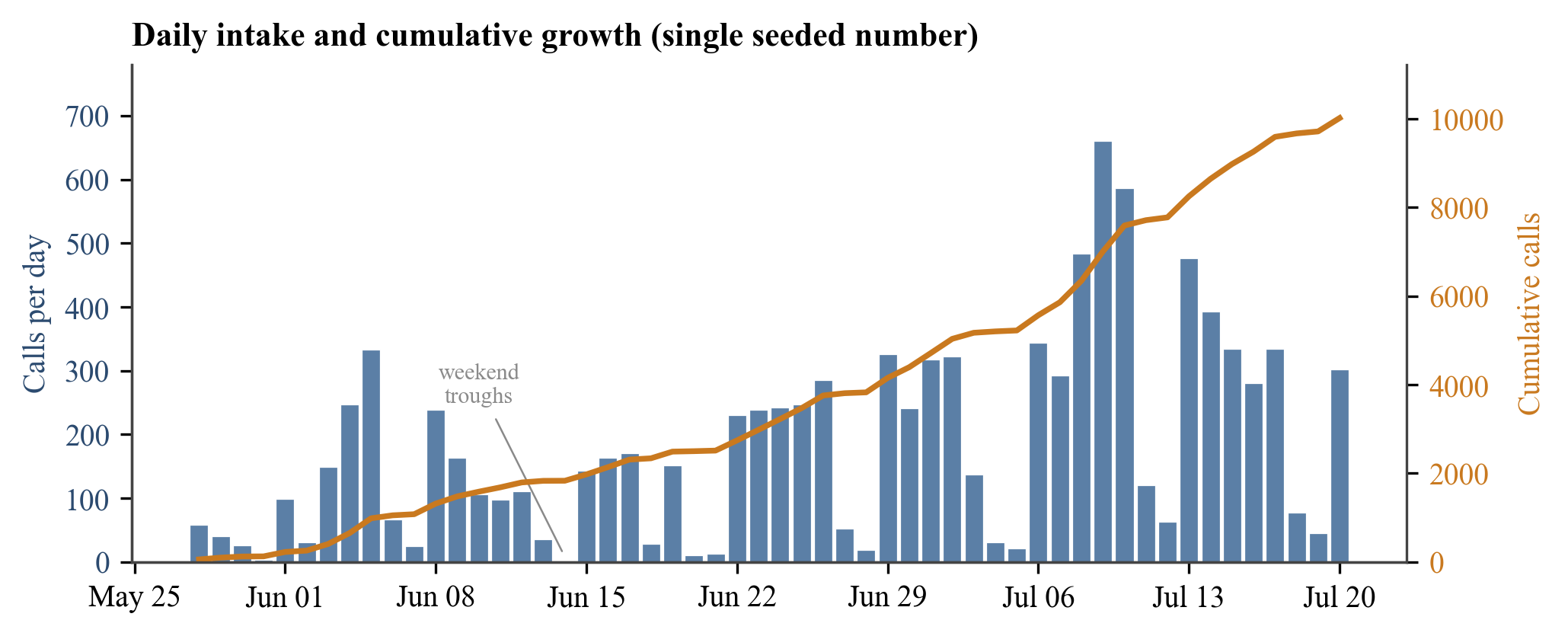}
\caption{Daily inbound call volume (bars, left axis) and cumulative total (line, right axis) over the 53-day window. Volume is weekly-cyclical---fraudulent lead generation operates during business hours, so intake craters on weekends---and trends upward as the seeded number propagates through resale lists.}
\label{fig:growth}
\end{figure}

This paper introduces a dataset that fills that gap and the system that produces it. The approach couples \textbf{active honeypot seeding} with a \textbf{conversational voice agent}: rather than waiting for calls to arrive, the system places dedicated numbers into the lead-generation and aggregator channels that fraud operations buy from, and then answers each inbound call with an adaptive persona engineered to keep the caller engaged long enough to reveal the full scam. Every call is recorded, transcribed at the level of individual turns, and passed through an automated labeling pipeline. From a single predominantly seeded line, the collection grew from roughly 127 calls in its first week to more than 2,500 in a peak week (Figure~\ref{fig:growth}), and the methodology is being extended to a fleet of distinct personas and numbers.

To our knowledge this is the \textbf{first large, publicly released corpus of real, multi-turn scam- and spam-call conversations collected autonomously by LLM voice agents}. It is distinct from robocall honeypots that answer calls but do not converse and do not release content \citep{robocalls,snorcall}, from LLM scam-baiting studies confined to email and text messaging \citep{sendaccount,aiintheloop}, and from corpora built on simulated or role-played dialogue \citep{botwars,puppeteer}. Our contributions are:
\begin{itemize}\setlength\itemsep{1pt}
\item \textbf{A collection methodology}: an active honeypot that couples lead-channel seeding with a low-latency LLM voice agent, turning scammer outreach itself into a continuous, self-refreshing data source.
\item \textbf{A corpus}: 10,015 real inbound scam and spam calls ($\approx$895 hours, 328,869 turns, 5,665 distinct callers) with transcripts, three-channel audio, latency telemetry, and multi-layer labels, including a human-corroborated holistic scam/spam/legitimate judgment; a curated de-identified subset is publicly released.
\item \textbf{Validation and benchmarks}: evidence that the calls are genuine and the engagement natural ($\approx$5\% bot-recognition), that labels align with human judgment (75\%), and that established detectors work on the corpus while synthetic-trained detectors collapse in precision on real traffic ($F_1$ 0.02--0.40).
\end{itemize}

The released corpus comprises \textbf{10,015 inbound calls collected between 28 May and 20 July 2026}, of which 6,601 contain two or more conversational turns. Table~\ref{tab:glance} summarizes the corpus. The remainder of the paper documents the collection system, the record structure, technical validation, and usage and ethics notes.

\section{Related Work}
\paragraph{Telephony honeypots and robocall measurement.} Passive honeypots assign unused numbers and characterize inbound traffic from metadata and audio \citep{phoneypot,robocalls}, and weak supervision has since scaled robocall \emph{content} analysis to hundreds of thousands of transcripts \citep{snorcall}. These systems answer mechanically but do not hold a conversation, target automated robocalls rather than live fraud operators, and do not release their transcripts; our corpus is conversational, two-way, and publicly released. Complementary threat-intelligence work establishes phone numbers as stable, high-value identifiers and studies how they propagate \citep{phonenum,callme,mobipot,sunshine,exploitphone}, and large-scale studies characterize specific fraud verticals such as technical-support scams \citep{dialone}.

\paragraph{Scam-baiting and defensive agents.} Keeping an operator engaged is itself valuable: the fixed-script Lenny bot occupies spammers \citep{lenny}, and a phone virtual assistant can screen robocalls by interrogating callers \citep{vamediated}. Recent LLM systems automate engagement, but overwhelmingly in the email and text-messaging setting \citep{sendaccount,aiintheloop}, or evaluate voice baiting against \emph{simulated} or role-played scammers rather than real inbound calls \citep{botwars,puppeteer,scambaitanalysis}. Our agent engages real scammers over voice at scale and, crucially, retains every call as released data.

\paragraph{Scam-detection data.} Detection models are typically trained on synthetic dialogue or on proprietary, unreleased call sets \citep{wheredowestand,teleantifraud}; the resulting detectors have not been evaluated on real, in-the-wild conversations. We show (Technical Validation) that this gap matters: detectors trained on published synthetic data collapse in precision when applied to our real calls.

\begin{table}[t]
\centering\small
\begin{tabular}{@{}p{0.56\columnwidth}p{0.36\columnwidth}@{}}
\toprule
\textbf{Property} & \textbf{Value}\\
\midrule
Collection window & 28 May -- 20 Jul 2026 (53 days)\\
Total inbound calls & 10,015\\
Substantive calls ($\geq$2 turns) & 6,601\\
Distinct originating numbers & 5,665\\
Recorded audio & $\approx$895 h (3-channel WAV)\\
Transcribed turns & 328,869\\
Turns per call (mean/med/max) & 50 / 17 / 1,341\\
Duration (med/mean/longest) & 2.2 min / 7.5 min / $\approx$1.8 h\\
Median agent reply latency & 1.17 s\\
Holistic labels (of 6,374) & 949 scam / 3,949 spam / 380 legit / 1,096 too brief\\
Reached a credential request & 710\\
\bottomrule
\end{tabular}
\caption{The corpus at a glance. Counts reflect real inbound traffic; development and self-test calls are excluded.}
\label{tab:glance}
\end{table}

\section{The Collection System}
\begin{figure}[t]
\centering
\includegraphics[width=\columnwidth]{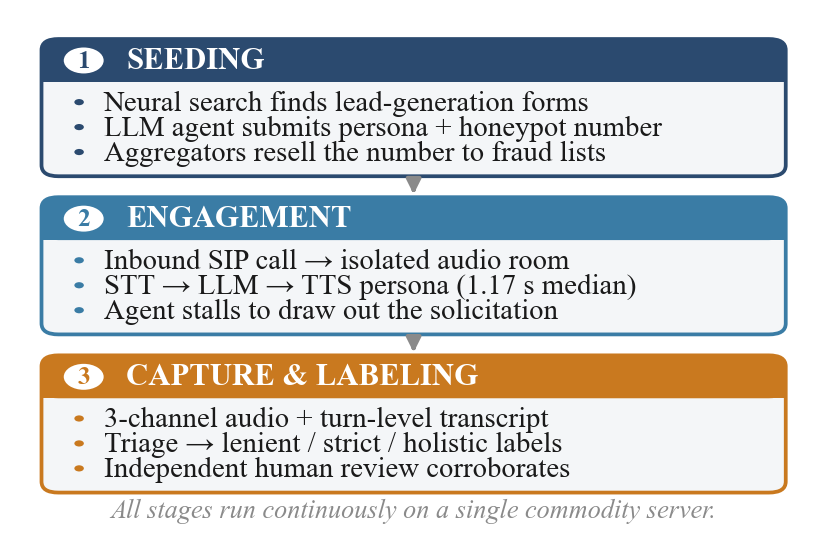}
\caption{The three-stage collection pipeline. Seeding manufactures inbound volume by placing a honeypot number into lead-generation channels; engagement answers each call with a low-latency persona agent that sustains the conversation; capture and labeling records every call and assigns the label layers released with the corpus.}
\label{fig:pipeline}
\end{figure}

The pipeline has three stages---\textbf{seeding} (generate inbound calls), \textbf{engagement} (answer and sustain the conversation), and \textbf{capture and labeling} (record and annotate)---which run continuously on a single commodity server (Figure~\ref{fig:pipeline}).

\subsection{Seeding}
The seeding stage manufactures inbound volume. A neural web-search service discovers lead-generation web forms---the ``request a quote'' and ``call me back'' pages for insurance, home security, auto warranties, debt relief, and similar verticals---whose submissions are aggregated and resold into the call lists that fraud operations purchase. A form-filling agent, driven by a language model that plans field-by-field entries, submits a consistent target identity together with a dedicated honeypot number through a pool of parallel headless-browser workers. Because aggregators resell submitted leads within hours, a seeded number begins receiving calls the same day. The target identity used for the bulk of the collection is a fictitious persona---an elderly, recently widowed retiree---chosen to present a high-value, plausible profile; the system is being extended to a fleet of personas that vary in age, gender, region, and disposition. That fraudulent lead buyers respond to seeding is confirmed directly within the conversations themselves (see Technical Validation).

\subsection{Conversational Engagement}
Inbound calls arrive over SIP into a real-time media server that places each caller into an isolated audio room. The production agent runs a cascaded speech pipeline: streaming speech-to-text transcribes the caller, a language model generates the reply, and neural text-to-speech renders it in the persona's voice. The persona prompt is engineered for \emph{engagement rather than refusal}: the agent stalls with clarifying questions instead of dead-ending the caller, accepts the premises the caller introduces, and expresses hesitation and emotion through inline paralinguistic cues (e.g., \emph{[nervous]}, \emph{[long pause]}) that the text-to-speech layer renders as audible affect. Sustaining natural conversation requires low latency; the median agent reply latency is 1.17 seconds, within the range of ordinary telephone conversation. A second agent variant uses a native speech-to-speech model; both share the persona design, and each call records which agent answered. The design goal is to maximize the fraudulent behavior elicited per call---in particular, to draw the caller past the pretext and into an explicit solicitation.

\subsection{Capture}
Every call is recorded to three synchronized WAV channels---agent audio, caller audio, and a mixed stereo track---and each conversational turn is written to a relational store with its transcript text, speaker role, timestamp, and a decomposition of response latency into voice-activity, model, and audio-synthesis components. Recording continues to the end of the call, so late-arriving content (such as a credential request after several minutes of pretext) is fully preserved.

\subsection{Automatic Labeling}
Labels are assigned in stages. A deterministic \textbf{triage} step removes non-substantive calls---IVR-only recordings, calls too brief to contain content, calls with no caller speech---without invoking a model. Substantive calls pass to a language-model \textbf{classifier} that assigns a lenient scam/not-scam judgment with a confidence score and behavioral signals drawn from a controlled vocabulary (e.g., \emph{medicare\_pretext}, \emph{impersonates\_insurance}, \emph{requested\_ssn}). A second, \textbf{strict} classifier marks a call as scam only when the caller actually solicited sensitive information---a Social Security number, Medicare or bank details, a payment card, or a date of birth---capturing the moment a lead-qualification call becomes an attempt at fraud; auxiliary passes label the call's opening move and ending. To avoid labeling a call before its late content has arrived, the strict classifier processes a call only after it has been inactive for a settling interval. A third pass assigns a single \textbf{holistic} judgment---scam, spam, or legitimate---that operationalizes the distinction between outright fraud and predatory-but-legal lead generation directly; it is prompted with the same rubric given to human reviewers and is corroborated against their judgments. Finally, a subset of calls is scored independently by a commercial voice-fraud classifier as an external check.

\section{Data Records}
The dataset comprises a relational metadata-and-transcript store and the underlying audio. The store holds two tables: \emph{sessions}, one row per call carrying identifiers, timing, duration, turn count, answering agent and persona, and every label layer (Table~\ref{tab:schema} lists the fields as released); and \emph{messages}, one row per turn with speaker role, transcript text, timestamp, and the latency decomposition for agent turns. Audio comprises 30,687 WAV files---three per call---totaling 614\,GB; the store itself is compact (98\,MB) and can be distributed independently for text-only use. On the lenient field a three-valued convention applies: true = scam, false = not, null = \emph{ignored} (removed in triage).

\begin{figure}[t]
\centering
\includegraphics[width=0.86\columnwidth]{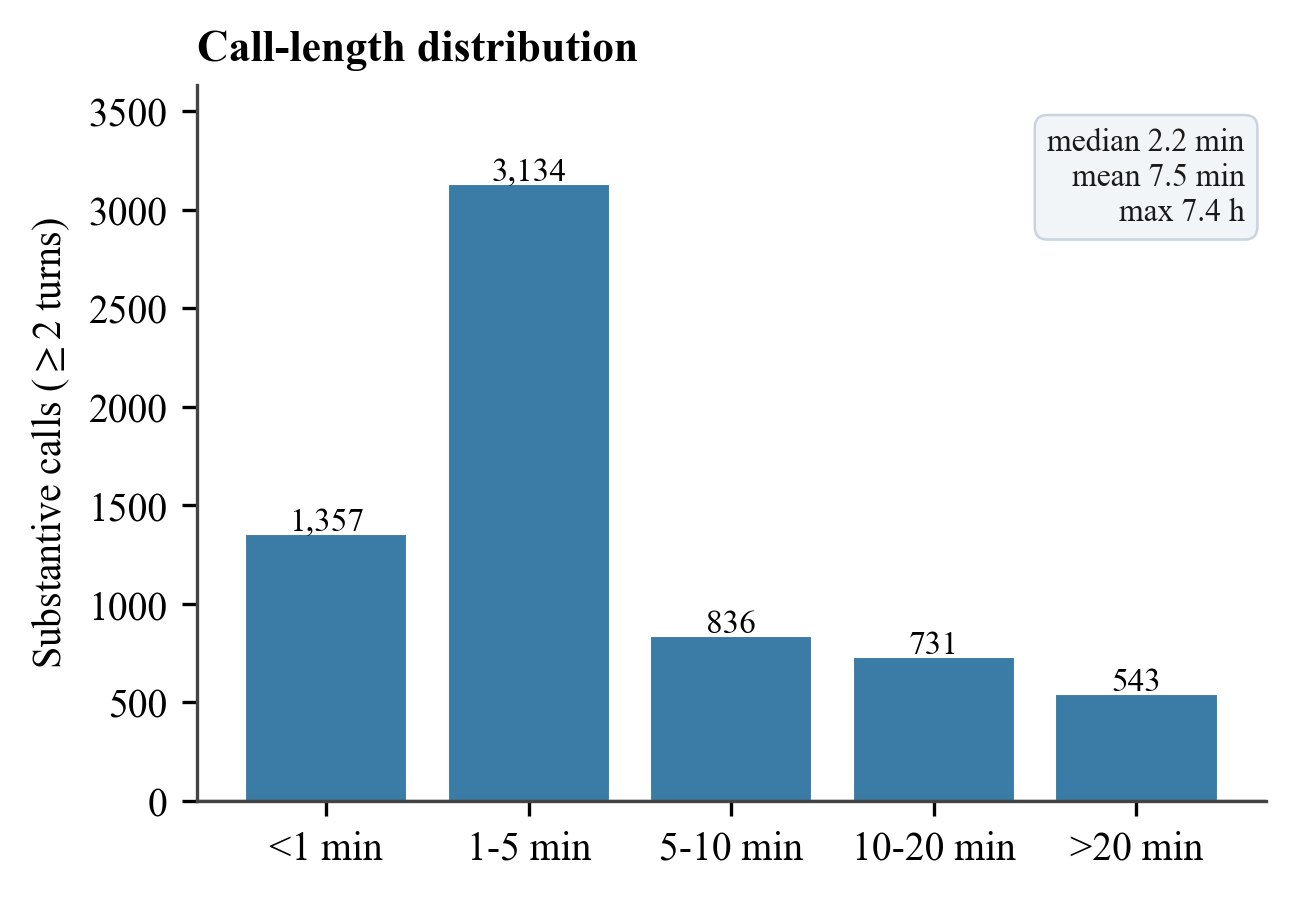}
\caption{Distribution of call length over the 6,601 substantive calls. The mass at one-to-five minutes reflects lead-qualification pitches; the tail beyond twenty minutes reflects prolonged engagements.}
\label{fig:duration}
\end{figure}

\begin{figure}[t]
\centering
\includegraphics[width=\columnwidth]{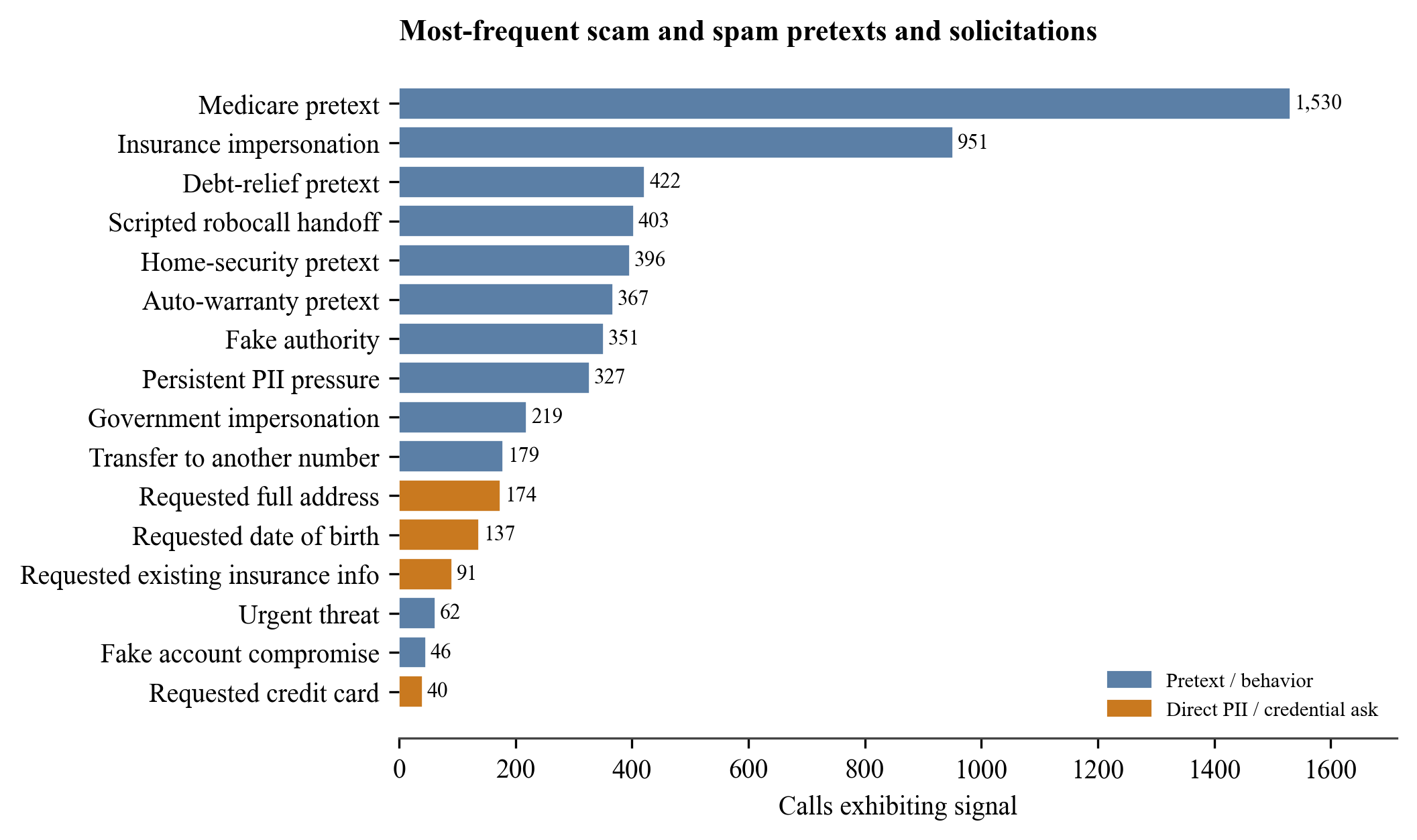}
\caption{Most-frequent behavioral signals assigned by the lenient classifier (calls may carry several). Blue: scam pretexts and tactics; orange: direct requests for personally identifiable or credential information.}
\label{fig:pretexts}
\end{figure}

The conversations are not monolithic. Most substantive calls run one to five minutes, but a long tail extends past twenty minutes and the longest genuine engagements exceed 1,200 turns over roughly 1.8 hours (Figure~\ref{fig:duration}). A single seeded persona already attracts a wide range of scam playbooks (Figure~\ref{fig:pretexts}): Medicare and insurance pretexts, home-security, auto-warranty, and debt-relief pitches, the government-, bank-, and technical-support impersonations characterized in prior studies of voice social engineering \citep{dialone}, and direct solicitations of addresses, dates of birth, and Social Security numbers.

\section{The Public Release}
The complete corpus---including three-channel audio and unredacted transcripts---is available under a data-use agreement. So the resource is usable without that step, we additionally release an openly downloadable, de-identified subset of \textbf{1,000 calls}: this section specifies what it contains, how those calls were selected, and how the subset departs from the corpus it is drawn from.

\subsection{Record Schema}
Each record is a self-contained JSON object combining per-call metadata, every label layer, and the full turn sequence, so no join is required; Table~\ref{tab:schema} gives the schema. A flat CSV of all fields except \texttt{turns} and \texttt{signals} accompanies the archive for call-level use. Categorical fields draw from closed vocabularies, so they serve directly as targets.

\begin{table*}[t]
\centering\small
\begin{tabular}{@{}llp{0.60\textwidth}@{}}
\toprule
\textbf{Field} & \textbf{Type} & \textbf{Description}\\
\midrule
\texttt{call\_id} & string & Pseudonymous call identifier.\\
\texttt{caller\_id} & string & Pseudonymous originating-number identifier; groups calls placed from the same number.\\
\texttt{persona} & string & Honeypot persona that answered the call (11 distinct personas).\\
\texttt{date} & date & Date of the call; time-of-day is withheld.\\
\texttt{duration\_s} & int & Call duration in seconds.\\
\texttt{n\_turns} & int & Number of transcribed conversational turns.\\
\texttt{label\_holistic} & enum & \emph{Primary label}: \texttt{scam} (fraud) or \texttt{spam} (predatory-but-legal sales).\\
\texttt{holistic\_confidence} & enum & Classifier confidence in the primary label (\texttt{high} / \texttt{low}).\\
\texttt{holistic\_impersonation} & bool & Caller impersonated a trusted institution.\\
\texttt{label\_scam} & bool & Strict label: the caller explicitly solicited sensitive credentials.\\
\texttt{label\_predatory} & bool & Lenient label: flagged predatory on pretext and tactics.\\
\texttt{opening\_type} & enum & Opening-strategy taxonomy label.\\
\texttt{ending\_type} & enum & Call-ending taxonomy label.\\
\texttt{asks} & list & Credential categories solicited (\texttt{ssn}, \texttt{medicare\_id}, \texttt{credit\_card}, \texttt{debit\_card}, \texttt{bank\_routing}).\\
\texttt{signals} & list & Behavioral-signal labels from the controlled vocabulary (Figure~\ref{fig:pretexts}).\\
\texttt{turns} & list & Ordered turns, each \{\texttt{speaker}: \texttt{caller}\,$|$\,\texttt{agent}, \texttt{text}\}; text is de-identified.\\
\bottomrule
\end{tabular}
\caption{Schema of a released record. Every label layer described in the Collection System is retained, so the three notions of ``scam''---lenient, strict, and holistic---remain separable by downstream users rather than being collapsed into a single flag.}
\label{tab:schema}
\end{table*}

\subsection{Composition and Selection Bias}
The subset pairs the 500 longest-engagement outright scams with the 500 longest-engagement predatory-spam calls, giving a balanced two-class resource of 183,491 turns spanning 28 May to 19 July 2026 (Table~\ref{tab:release}). Legitimate calls are excluded entirely, and audio is withheld. Eleven personas appear, though the collection remains dominated by the original seeded persona (573 of 1,000 calls).

Selection was deliberately \emph{not} random, and the consequence must be stated plainly: released calls are far longer than typical corpus traffic---median 19.8 minutes against 2.2 corpus-wide, with no released call shorter than 4.8 minutes. The subset is a \textbf{high-engagement stratum}, chosen because long calls contain complete scam scripts and are more informative per record. It should not be used to estimate prevalence or base rates; Table~\ref{tab:glance} and Figures~\ref{fig:duration}--\ref{fig:funnel}, computed over the full corpus, remain the reference for those. Within the subset the classes also differ in length (median 14.7 minutes for scams against 23.0 for spam), so length is not a proxy for fraud.

\begin{table}[t]
\centering\small
\begin{tabular}{@{}p{0.56\columnwidth}p{0.36\columnwidth}@{}}
\toprule
\textbf{Property} & \textbf{Value}\\
\midrule
Released calls & 1,000 (500 scam / 500 spam)\\
Window & 28 May -- 19 Jul 2026\\
Transcribed turns & 183,491\\
Turns per call (med/mean) & 148 / 184\\
Duration (min/med/max) & 4.8 min / 19.8 min / 4.0 h\\
Distinct callers & 733\\
Repeat callers ($>$1 call) & 153 (420 calls)\\
Calls with a credential ask & 519 (52\%)\\
High-confidence primary label & 946 (95\%)\\
\bottomrule
\end{tabular}
\caption{The public release at a glance. Because the subset is selected for engagement length, these distributions differ by construction from the full corpus in Table~\ref{tab:glance}.}
\label{tab:release}
\end{table}

\subsection{What the Label Layers Separate}
Retaining three label layers lets users choose an operational definition rather than inherit ours, and the layers demonstrably disagree. Among released scams, 64\% reached an explicit credential request under the strict classifier against 40\% of spam calls---a credential ask is common in predatory sales and is therefore insufficient on its own to define fraud. Institutional impersonation separates the classes far more sharply (25\% against 1\%), while the lenient layer flags most of both (74\% and 84\%), functioning as a sensitive screen rather than a decision rule.

Where a credential ask occurs, Social Security numbers dominate (327 calls, 33\%), followed by payment cards (188), bank routing details (136), Medicare identifiers (72), and debit cards (61), often in combination. The ending taxonomy records how attempts resolved: operators most often punted the call onward (420) or the line dropped (242), 110 gave up, and in 56 the operator proceeded as though the solicitation had succeeded---against wholly fictitious information. A further 63 calls (6\%) ended with the caller apparently recognizing an automated system.

\subsection{Suggested Evaluation Protocol}
The subset's caller structure has direct methodological consequences. Its 1,000 calls originate from 733 distinct numbers, 153 of which appear more than once (up to 12 times), so \textbf{420 of the 1,000 calls---42\%---come from a caller that also appears elsewhere in the subset}. A random train/test split therefore places the same operator, reciting the same script, on both sides of the partition, and inflates apparent performance by rewarding memorization of caller-specific phrasing rather than generalization to unseen operations. We accordingly ship caller-disjoint fold assignments with the archive and report all benchmarks below under that protocol; users comparing against published numbers should confirm which regime those numbers used.

\section{Technical Validation}
We validate three properties: that the calls are genuine and the engagement natural; that seeding, not chance, produces the traffic; and that the labels are meaningful.

\subsection{The Calls Are Real, the Engagement Natural}
Three lines of evidence indicate that callers treat the honeypot as a live human target. First, the agent responds quickly enough to sustain ordinary conversation: median reply latency 1.17\,s (mean 1.55\,s) over 129,425 agent turns. Second, conversations reach substantial depth---a mean of 50 turns per substantive call, 543 calls exceeding twenty minutes---far beyond the immediate hang-ups that dominate passive honeypots \citep{phoneypot,robocalls}. Third, and most directly, callers act on the seeded identity: in the excerpt below, the caller recites the persona's seeded details back as the reason for the call, then escalates to an explicit request for a Social Security number---exactly the behavior the pipeline is designed to elicit and record.

\begin{mdframed}[backgroundcolor=black!4,linewidth=0.5pt,innertopmargin=4pt,innerbottommargin=4pt]
\small
\textbf{Caller:} \emph{You had put in a site builder notification\ldots\ it says your name, Edna Whitcomb. It says your ZIP code is 3-4-2-3-9\ldots\ gave me the phone number that I just called\ldots\ you are looking for homeowners [and] umbrella, so that's why I'm calling.}

\textbf{Agent:} \emph{[confused] Oh, I see. Yes, that sounds right\ldots\ So what exactly does this ``umbrella'' mean for my homeowners?}

\textbf{Caller:} \emph{In order to do that\ldots\ I will need to have the following information. Your name, your date of birth, your Social Security number\ldots\ the physical address\ldots\ your mailing address.}

\textbf{Agent:} \emph{[nervous] Oh, my Social Security number? I don't usually give that out over the phone, you know. Is there another way to get a quote without that, dear? My son always tells me not to give out that kind of information.}
\end{mdframed}
\noindent The agent is rarely unmasked: of the 6,601 substantive calls, only 339 ($\approx$5\%) end with the caller apparently recognizing an automated system---closely matching the $\approx$5\% bot-recognition rate reported for the Lenny scam-baiting bot \citep{lenny}.

\subsection{Seeding Drives the Traffic}
The honeypot numbers are freshly provisioned and never used by real people, so all inbound traffic is unsolicited; yet from a predominantly single seeded line, weekly intake rose from about 127 calls to more than 2,500 in a peak week (Figure~\ref{fig:growth}), drawn from 5,665 distinct originating numbers. The pronounced weekly cycle---near-zero volume on weekends, recovering each Monday---matches the business-hours operation of commercial lead generation and is itself evidence that the callers are organized operations working seeded lists, not random misdials.

\subsection{The Labels Are Meaningful}
\begin{figure}[t]
\centering
\includegraphics[width=\columnwidth]{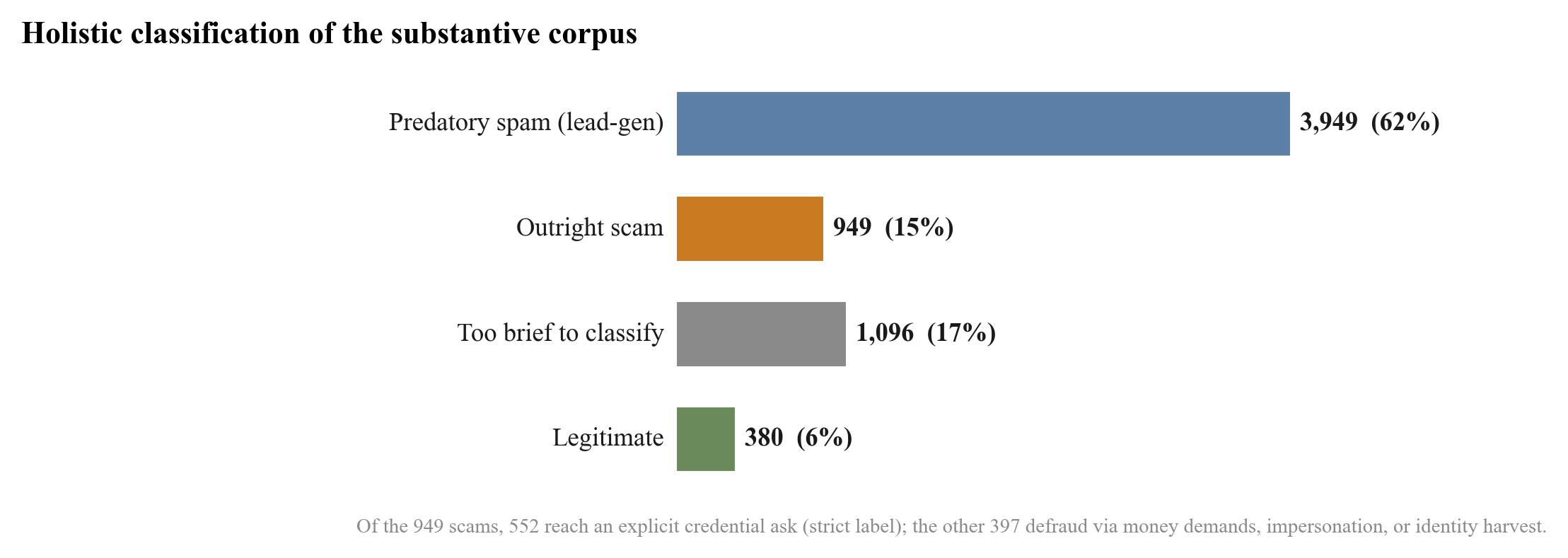}
\caption{Holistic classification of the substantive corpus. The majority is predatory spam; about one in seven is an outright scam. Of the 949 scams, 552 reach an explicit credential request; the remainder defraud through payment demands, impersonation, or identity harvesting.}
\label{fig:funnel}
\end{figure}

\begin{figure}[t]
\centering
\includegraphics[width=\columnwidth]{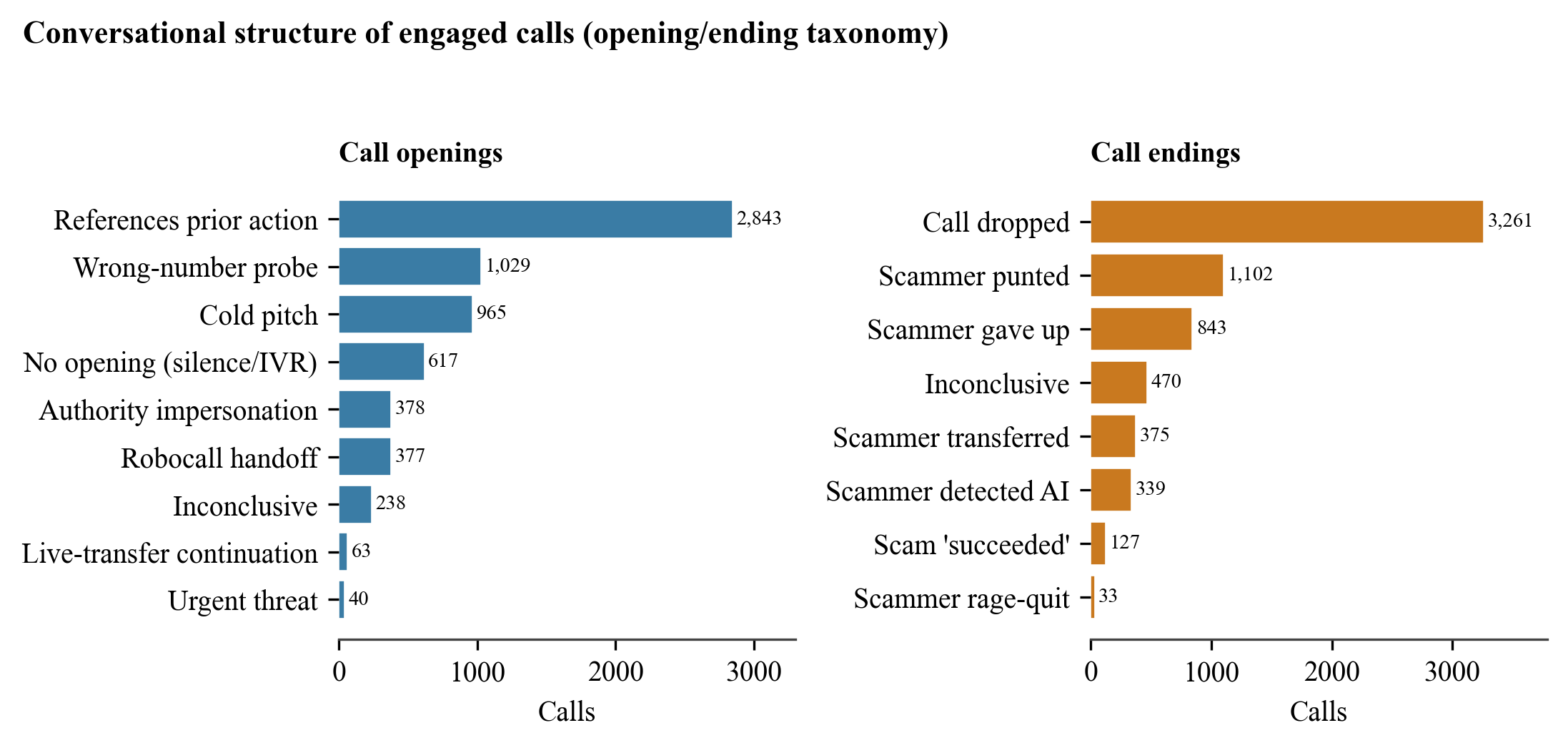}
\caption{Conversational structure of engaged calls under the opening (left) and ending (right) taxonomies---a coarse behavioral fingerprint beyond the binary scam label.}
\label{fig:structure}
\end{figure}

The two-stage design separates deterministic removal of non-substantive calls from model judgment on the remainder, and the strict classifier settles each call before labeling so late credential requests are not missed. The three classifiers capture complementary notions of ``scam''. Of 6,601 substantive calls, 6,374 have settled: the holistic classifier labels 949 ($\approx$1 in 7) outright scams and 3,949 predatory-but-legal spam (Figure~\ref{fig:funnel}); the lenient classifier flags 4,404 on pretext and tactics; the strict classifier isolates the 710 that reached an explicit request for sensitive information. A commercial voice-fraud classifier scored a subset as an independent check, agreeing on a clear majority with disagreements concentrated on short, ambiguous calls. Pipeline labels are best understood as high-quality automatic (``silver'') annotations rather than adjudicated ground truth. The opening and ending taxonomies (Figure~\ref{fig:structure}) describe each conversation orthogonally---separating calls that open by referencing a prior action from cold pitches and wrong-number probes, and dropped lines from operators who gave up or transferred.

\subsection{Human Review Corroborates the Labels}
Six independent reviewers each labeled a stratified sample of calls---60 in total---under a three-way rubric: outright \emph{scam} (a deceptive attempt to obtain sensitive information or money, typically through institutional impersonation or a fabricated pretext), predatory-but-legal \emph{spam} (a sales or lead-qualification pitch), or \emph{legitimate}. On the binary scam/not-scam decision the strict credential-ask label agreed with reviewers on 67\% of calls; the classifier prompted with the reviewers' own rubric---released as the holistic label---raised agreement to 75\%. Residual disagreement fell almost entirely on the boundary between outright scam and predatory spam: identity-harvesting debt-relief, payday-loan, and ``soft credit pull'' pitches on which the reviewers themselves split. Manual inspection found the classifier often the more consistent annotator---one reviewer, for example, marked as ordinary spam a call that solicited both a Medicare identifier and a Social Security number. We therefore treat the scam/spam frontier as genuinely graded rather than as classifier error, release the human annotations alongside the corpus, and are enlarging the annotated subset. Consistent with its broader scope, the holistic scam label is not a subset of the strict label: it additionally captures fraud pursued through payment demands, impersonation, and bulk identity harvesting that never reaches a strict-category ask, while a minority of calls that do request a payment card or Medicare number are judged spam---a genuine sale being closed.

\subsection{Established Detection Methods Are Effective on the Corpus}
\begin{figure}[t]
\centering
\includegraphics[width=\columnwidth]{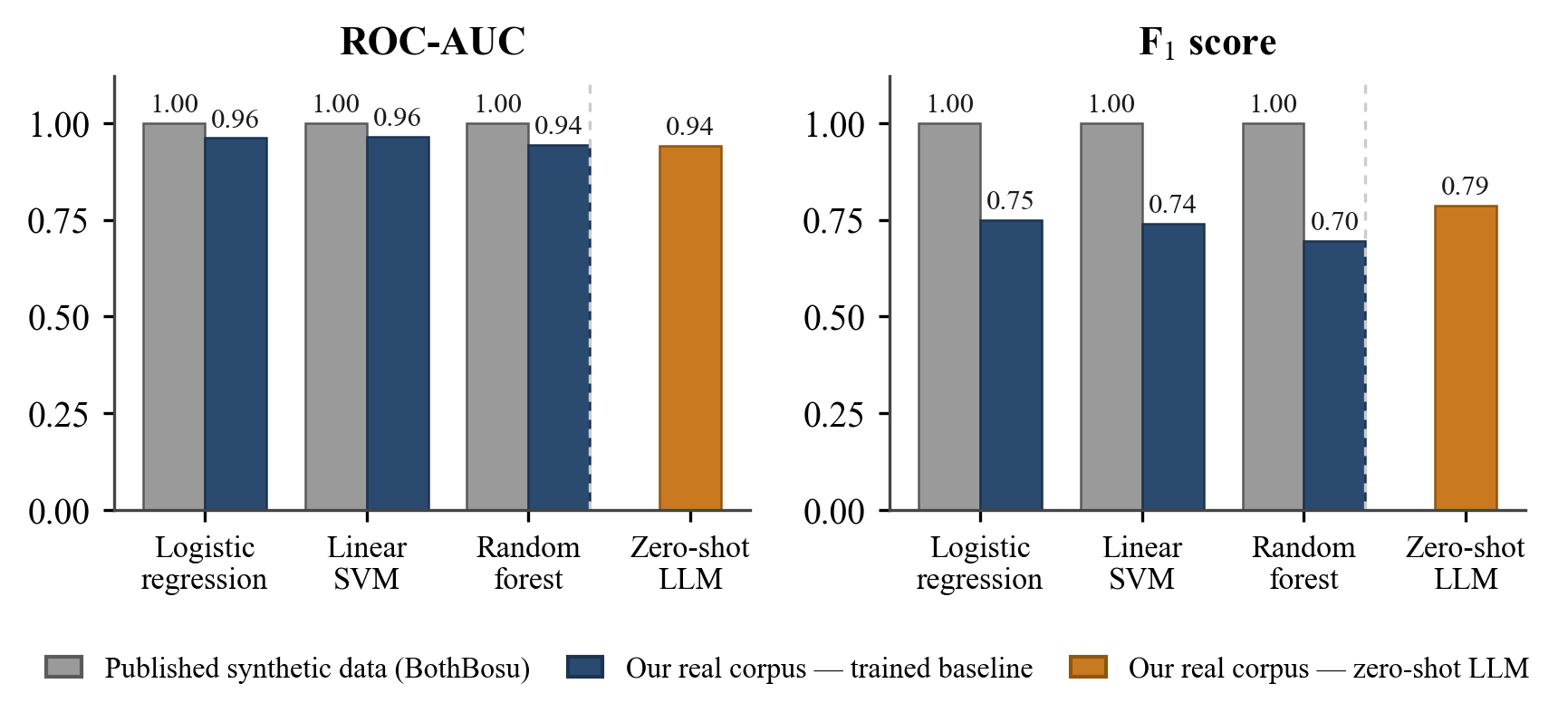}
\caption{Established methods benchmarked on the corpus. TF-IDF baselines \citep{wheredowestand} reproduced on their synthetic data (grey, ROC-AUC/$F_1$ = 1.00), then applied unchanged to our real corpus (navy; caller-disjoint CV); a zero-shot LLM (orange) shown for comparison. Left: ROC-AUC; right: $F_1$ at a 15\% base rate.}
\label{fig:bench}
\end{figure}

A data resource is validated in part by whether standard methods can be trained and evaluated on it. We take the classical text-classification baselines that a recent survey of phone-scam detection reports at near-perfect accuracy on synthetic corpora \citep{wheredowestand}---TF-IDF n-gram features with logistic regression, a linear SVM, and a random forest---and first reproduce them on that published data (the BothBosu scam/non-scam dialogue sets, 3,840 balanced dialogues), where all three reach ROC-AUC and $F_1$ of 1.00, confirming the implementation. Applied to our corpus---each transcript labeled by the holistic classifier, caller-disjoint five-fold cross-validation---the same methods separate outright scams from surrounding spam and legitimate calls with ROC-AUC 0.94--0.96 and $F_1$ 0.70--0.75 at a 15\% positive base rate (Figure~\ref{fig:bench}). A training-free zero-shot LLM detector (Gemini 2.5 Flash) given the same holistic rubric reaches ROC-AUC 0.94 and $F_1$ 0.79 (precision 0.73 at 0.85 recall)---on par with the trained baselines. Two caveats frame this honestly. Whole-call detection is comparatively easy because a full transcript usually contains the pretext and request in the clear; the harder task is predicting escalation from only a caller's opening turns, a setting the corpus equally supports \citep{loralit}. And a detector trained on the published synthetic corpus and applied unchanged to our real calls preserves only part of its ranking ability (ROC-AUC 0.75--0.78) but collapses in precision---$F_1$ falls to 0.02--0.40---direct evidence that models tuned on synthetic data do not transfer to live conditions, and that a corpus drawn from real traffic is needed to evaluate them. The caller-disjoint split assignments are provided with the released data so these evaluations can be reproduced.

\section{Usage Notes}
\subsection{Intended Uses}
The corpus supports research on telephone fraud and its mitigation: training and benchmarking scam- and vishing-detection models on realistic dialogue; analyzing social-engineering tactics and the escalation from pretext to credential solicitation; building taxonomies of scam scripts; and developing defensive conversational agents that detect, delay, or disrupt fraud in real time \citep{sendaccount,aiintheloop,warnedme}. The turn-level latency telemetry additionally supports research on real-time spoken-dialogue systems.

\subsection{Limitations}
The collection is English-language and U.S.-centric, and to date the large majority of calls were answered by a single persona on a single number, so persona-conditioned analyses are limited until the multi-persona fleet accumulates volume. The class distribution is imbalanced: only a minority of calls reach an explicit credential request, and much of the traffic is predatory-but-legal lead generation rather than outright fraud. All labels are automatic and should be treated as silver annotations; transcripts contain recognition errors, and the agent's synthesized speech may introduce artifacts absent from human-to-human calls. The human-annotated subset is modest and singly annotated, so it corroborates the automatic labels rather than yielding a precise inter-annotator agreement estimate. Finally, because the agent actively engages callers, the dataset reflects scammer behavior in response to an engaged, plausible target, not a neutral sample of all inbound scam traffic.

\subsection{Ethics and Responsible Use}
The collection is designed to avoid harm to third parties. The honeypot answers only \emph{inbound} calls to dedicated numbers that are never advertised to, or used by, real people; it initiates no outbound calls, performs no intervention with potential victims, and collects no data about them. The only parties recorded are operators who placed an unsolicited call to a seeded line in order to defraud or to sell to someone they believed to be a consumer. Recording is confined to these calls: the research team owns and operates every honeypot line and is therefore a party to each recorded call, satisfying one-party-consent requirements; in jurisdictions that require all-party consent, the recordings capture only unsolicited callers to researcher-controlled numbers, are used solely for security research, and are released only in de-identified form---the framing established by prior telephony-honeypot studies that engage and record callers \citep{robocalls,blacklist}. The persona's presentation as a human target is a minimal, proportionate deception directed exclusively at parties actively attempting fraud; no non-fraudulent party is knowingly deceived. These choices follow the beneficence and justice balancing of the Menlo Report \citep{menlo}, under whose principles the study was conducted; because it records only unsolicited callers to researcher-operated lines and involves no human-subject victims, the work does not constitute human-subjects research.

Seeding places a fictitious identity---not any real person's information---into lead-generation channels, so it exposes no individual and diverts fraud attempts toward the honeypot rather than a genuine consumer. A small fraction of inbound calls are legitimate; these are labeled as such and \emph{excluded} from the public release, which contains only scam and spam calls. The release is de-identified---originating numbers pseudonymized, and spoken names, phone numbers, addresses, and organization names redacted to placeholders---and documented with a datasheet \citep{fair,datasheets}. Audio, which carries voiceprint re-identification risk, is withheld. Users must not employ the corpus to build systems that defraud, harass, or deceive, must not attempt to re-identify any party, and must comply with applicable call-recording and data-protection law.

\section{Code and Data Availability}
The evaluation and analysis code reproducing the reported benchmarks and release statistics is released with the data; the collection pipeline is withheld on dual-use grounds and available from the authors. The public release is distributed under a CC BY-NC license with a datasheet; the full corpus and audio are available under a data-use agreement. A persistent identifier will be assigned on publication.

The canonical, maintained location for the released resources associated with this work is the
Scam.AI Research portal, \url{https://scam.ai/research}~\cite{scamai_research}; document- and
PDF-forgery resources are additionally maintained at Trustable, \url{https://trustable.ai}~\cite{trustable_ai}.
As noted above, the full corpus and audio are not part of the open release and remain available under a
data-use agreement; requests and further details are handled through the Scam.AI Research portal, which
supersedes any prior ad hoc distribution.

\bibliography{refs}

\end{document}